\documentclass[letterpaper,11pt]{article}

\usepackage{graphicx}
\usepackage{amsmath}
\usepackage{amsfonts}
\usepackage{amssymb}
\usepackage{graphicx}
\usepackage{mathrsfs}
\usepackage{siunitx}
\usepackage{array}
\usepackage{natbib}
\usepackage{colortbl}
\usepackage{caption}
\begin{document}

\title{Homotopy Sampling, with an Application to Particle Filters}
\author{J. M. Restrepo \footnote{omputer Science and Mathematics Division, Oak Ridge National Laboratory, Oak Ridge, TN 37831, USA} \footnote{Computer Science and Mathematics Division, Oak Ridge National Laboratory, Oak Ridge, TN 37831, USA},  J. M. Ram{\'i}rez\footnote{Escuela de Matem{\'a}ticas, Universidad the Colombia, Sede Medell{\'i}n,  Medell{\'i}n, Colombia}}

\maketitle

\begin{abstract}
We propose a homotopy sampling procedure, loosely based on importance sampling.  Starting from a known probability distribution, the homotopy procedure generates the unknown normalization of a target distribution. 

In the context of  stationary distributions that are associated with physical systems the method is an alternative way to  estimate an unknown microcanonical ensemble.
The process is iterative and also generates  samples from the target distribution.

In practice,  the homotopy procedure does not circumvent using  sample averages
in the estimation of the normalization constant. 
The error in the procedure depends on the errors incurred in sample averaging and the number of stages used in the 
computational implementation of the process.  However, we show that it is possible to exchange the number of homotopy stages and the total number of samples needed at each stage in order to enhance the computational efficiency of the implemented algorithm. Estimates of the error as a function of stages and sample averages are derived. These could  guide computational efficiency decisions on how the calculation would be mapped to a given computer  architecture. 
 
Consideration is given to how the procedure can be adapted to Bayesian estimation problems, both stationary and non-stationary. Emphasis is placed on the non-stationary problems, and in particular,  on a sequential estimation technique known as particle filtering. It is shown that a modification of the particle filter framework to include the homotopy process can improve the
computational robustness of particle filters. The homotopy process can 
 ameliorate particle filter collapse, a common challenge to using particle filters when the sample dimension is small compared with the state space dimensions.

\end{abstract}

\section{Introduction}
Homotopy sampling, or path sampling \cite{JSliu,mchandbook}, 
are names given to  a  broad collection of  methods that rely on analytical continuation (see
\cite{continuationbook}).
The idea is to construct a function $Z(s)$ that varies continuously with the homotopy parameter $s$. The parameter varies between $s_0$ and $s_1$. Knowing $Z(s_0)$ the homotopy procedure obtains the unknown  target $Z(s_1)$.

Here we explore the application of a specific homotopy, which is largely inspired by importance sampling \cite{JSliu}.  Generalizations of the homotopy process will be 	developed as well.
(In what follows we will denote a probability distribution function (pdf) $f(x)$  as {\it proper} if  $\int f(x) \, dx = 1$. Otherwise, we will say $f(x)$ is an improper pdf). 
The goal of the proposed homotopy is to obtain an estimate of the normalization constant  $Z_1:= \int q(x)\, dx$, for the target distribution $q(x)/Z_1$, starting from
the known constant $Z_0$ associated with the proper distribution
$p(x)/Z_0$. In the calculation of $Z_1$ we also obtain samples for the proper target distribution $q(x)/Z_1$.

  When the iterative process underlying the homotopy for finding $Z_1$is discretized in the parameter $s$ we will obtain a computational  
  algorithm that estimates $Z_1$. We connote the discretized  algorithm as the 
  {\it homotopy schedule}.   The homotopy iteration process and its approximation  will be described in Section \ref{sec.homot}.  In the practical application of the homotopy procedure  sample averaging will be used as a way to circumvent the challenge of computing the integrals analytically.
  Because of this, and the iterative nature of the schedule, errors in the estimate of the normalization constant compound and thus depend on the number of samples $N$ used at each homotopy stage, and the number of stages, $M$. Using 
   the central limit theorem we will derive the overall error of the homotopy 
   schedule, and how it depends on $M$ and $N$, as well as on $p(x)$. This estimate will appear in Section \ref{sec.error}. The elliptic interdependence of the error on the number of homotopy stages $M$ and the number of sample averages $N$ per stage leads to the possibility of exploiting these dependences in achieving computational efficiency by taking advantage of certain computer architectures. This will constitute  a central practical result in this paper. 
  In Section \ref{sec.complexity} we demonstrate numerically that the error estimate captures qualitatively the $N$, $M$ dependence of the error. 
   
 We will also consider how the homotopy process may be applied to finding the evidence  in Bayesian estimation. Since obtaining the normalization constant for a posterior distribution 
 involves more than one distribution, in Section \ref{sec.bayes} we will show a variety of different ways in which the homotopy procedure can be implemented. 
 A key finding is that when either the prior or the likelihood are proper distributions and either of these are used as a proposal for $p(x)$ (with the posterior assuming the role of the target distribution $q(x)$), there is a homotopy process that 
 removes the requirement that the support of $p(x)$ be larger than $q(x)$, which is otherwise required in the estimate of the less complex $q(x)$ cases considered in 
 Section \ref{sec.homot}.

 In Section \ref{sec.assimilation} we take up the non-stationary Bayesian case. 
 The focus is on Bayesian estimation of  noisy time dependent state variables, conditioned  on observations. 
 The particle filter  \cite{sarkkabook} is a sequential sampler that is used to find moments from a time dependent conditional distribution. Particle filters do not require 
 that we know the evidence in order to be applied in sampling the posterior distribution; the weights associated with the particles are normalized at each time step. However, this self-normalization procedure can lead to a concentration of weight among very few particles. This, in turn,  can lead to a serious loss of statistical significance and poor estimation and sometimes to `filter collapse'.
 This is  a particularly common problem 
 in the application of particle filters when the number of particles used is small, as compared to the dimension of the state variable. 
  Conventional particle filter 
 implementations often use some form of resampling in order to 
 ameliorate  this tendency toward `filter collapse' \cite{bickel}.  We will show by numerical means that the homotopy process can be used within the particle filter process to produce a more detailed probabilistic description of the state variables, in addition to mitigating filter collapse.
 
Section \ref{sec.conclusions} presents a summary of the method, our findings, and suggested avenues  for further inquiry.

\section{The Homotopy Proposal}
\label{sec.homot}

Let $x \in \mathbb{R}^{K}$ be the $K$-dimensional sample space and $q(x)$ an improper distribution which we will refer to as the {\it target} distribution. Associated with this distribution is the unknown constant $Z_1$ such that 
\[
Z_1= \int q(x) dx.
\]
We propose to use a homotopy procedure to  find $Z_1$. 
We will assume that we know the constant $Z_0$
  and the improper pdf $p(x)$, such that
$Z_0 = \int p(x) dx$.  Further, we will require that $p(x)$ be chosen so that  
\begin{equation}\label{Hyp_pq}
    \sup_{x \in \mathbb{R}}\frac{q(x)}{p(x)} < \infty
\end{equation}
 Let
   \begin{equation}
  \theta_s(x):=\frac{q^s(x)p^{(1-s)}(x)}{Z_s},
\label{thetaeq}
  \end{equation}
  where 
 \begin{equation}
 Z_s=\int q^s(x)p^{(1-s)}(x) dx
 \label{zeq}
 \end{equation}
  is a constant, and $s\in [0,1]$. We note that 
  \begin{equation}
 \theta_0 =p(x)/Z_0,  \theta_1  =q(x)/Z_1.
  \label{ends}
  \end{equation}
From (\ref{thetaeq}), (\ref{zeq})  and (\ref{ends})  it is surmised that 
\begin{equation}
 \ln\big(\theta_s(x)\big)=s\ln\big(q(x)\big)+(1-s)\ln\big(p(x)\big)-\ln Z_s, \qquad 0 \le s \le 1.
 \label{homot}
 \end{equation}
Further, we note that 
    \begin{equation}
    \frac{Z_{s+\epsilon}}{Z_s}=\frac{1}{Z_s} \int\bigg(\frac{q(x)}{p(x)}\bigg)^\epsilon p^{(1-s)}(x)q^s(x)dx :=\bigg\langle\bigg(\frac{q(x)}{p(x)}\bigg)^\epsilon\bigg\rangle_s.
    \label{zzs}
    \end{equation}
    The notation $\langle \cdot \rangle_s$ denotes the {\it expectation  with respect to the homotopy density} 
    $\theta_s(x)/Z_s$.
When  $\langle (q(x)/p(x))^\epsilon \rangle_s$  can be performed analytically, for $s \in [0,1]$, it is clear
that we obtain a function $Z_s$ whose initial value is known, and its target $Z_1$ is the desired
normalization constant. In practice this integral cannot be found analytically 
(and it would be of little practical utility since the integral for $s=1$ is equal to the desired constant $Z_1$). Hence, an iterative process is proposed that will allow us to find $Z_1$ or its approximation. 
 
Note that in the limit of $\epsilon \rightarrow 0$, we obtain a differential equation for $Z_s$, {\it viz},
 \[
\frac{d Z_s}{ds} = \int \log \left(\frac{q}{p} \right) q^s p^{1-s} dx:= \bigg \langle \log \left(\frac{q}{p} \right) \bigg \rangle_s   Z_s.
\]

\subsection{The Homotopy Schedule}

Let $s_m := m \epsilon$, $m=1,...,M$, and $\epsilon = 1/M$ (the intervals need not be equal as explained in Section \ref{sec.general}, but for now we assume they are).
We can then write $Z_1/Z_0$ using (\ref{zzs}) as the expanded product of fractions: 
\begin{align*}
\frac{Z_1}{Z_0}&=\frac{Z_\epsilon}{Z_0} \cdot \frac{Z_{2 \epsilon}}{Z_\epsilon} \cdots \frac{Z_1}{Z_{(M-1) \epsilon}} \\
  &=\prod_{m=1}^{M}\frac{Z_{m \epsilon}}{Z_{(m-1) \epsilon}}
  =\prod_{m=1}^{M}  \bigg\langle \bigg(\frac{q(x)}{p(x)}\bigg)^{\epsilon}\bigg\rangle_{(m-1) \epsilon}.
  \end{align*}
Taking a natural logarithm of each side we obtain
 \begin{equation}
 \ln\bigg(\frac{Z_1}{Z_0}\bigg)=\sum_{m=1}^{M}
  \ln \bigg\langle\bigg(\frac{q(x)}{p(x)}\bigg)^{\epsilon}\bigg\rangle_{(m-1)\epsilon}.
 \label{logeq}
 \end{equation}
 The equality in expression (\ref{zzs}) and (\ref{logeq})  are satisfied for any $0<\epsilon \le 1$, if the necessary expectation calculation can be done exactly. 
 However,  in practice  the requisite expectation calculations  need to be approximated by a finite sample average. In that case the homotopy schedule will yield an approximation $\bar Z_1$ that will depend on the number of samples $N$ used and the number of homotopy stages $M=1/\epsilon$.
 
 When the schedule is exact, the succinct summary of the procedure is as follows:
 Defining 
\begin{equation}
     w_{\epsilon}(x) := \left(\frac{q(x)}{p(x)}\right)^\epsilon
     \label{Def_eps}
\end{equation}
and
  \begin{equation}
 \mu_{m \epsilon}:=  \bigg\langle w_{\epsilon}\bigg\rangle_{m \epsilon}.
 \label{mu}
 \end{equation}
 Then 
 \begin{equation}
 Z_{(m+1)\epsilon} = Z_{m\epsilon} \mu_{m \epsilon}, \quad m \in [0,M-1],
 \label{zs}
 \end{equation}
 where $M=1,2,..$ and $M \epsilon=1$
 and
 \[
 Z_1 = \prod_{m=0}^{M-1} \mu_{m \epsilon}.
 \]
Presuming the expectation calculations are not exact and are instead approximated by an $N$ term sample average, then
  \begin{equation}
  	\ln\bigg(\frac{Z_1}{Z_0}\bigg)\approx\sum_{m=1}^{M}\ln\Bigg(\frac{1}{N}\sum_{n=1}^N \bigg(\frac{q(X(n)_{(m-1)})}{p(X(n)_{(m-1)})}\bigg)^\epsilon \Bigg), \label{Eq_lnZ1Z0}
  \end{equation}
 where  $X(n)_{(m-1)}$ is the $n^{th}$ sample from the (precomputed)  $(m-1)^{th}$ distribution
 \[
\theta_{(m-1)\epsilon}(x) = \frac{1}{Z_{(m-1) \epsilon}} q^{(m-1)\epsilon}(x)p^{1-(m-1)\epsilon}(x).
 \]
 \medskip
 The sample average approximation of $\mu_{m}$ on the $m$ lattice shall be denoted as
 \begin{equation}
 \bar \mu_{m \epsilon} := \frac{1}{N} \sum_{n=1}^N \left( \frac{q(X(n)_m)}{p(X(n)_m)} \right)^\epsilon.
\label{barmu}
 \end{equation}
 Then the homotopy schedule  (using sample averages) is given by
\begin{equation}
\bar{Z}_{(m+1) \epsilon} = \bar{Z}_{m\epsilon} \, \bar \mu_{m \epsilon},
\label{Eq_forwardDis}
\end{equation}
and approximation to $Z_1$, correspondingly, would be obtained as
\[
 \bar Z_1 = \prod_{m=0}^{M-1} \bar \mu_{m \epsilon}.
 \]

\subsection{Example Calculation}
\label{example}
Consider the estimation of $Z_1= \int_{-\infty}^\infty q(x) \, dx$, where
\[
q(x)=e^{-\frac{(x-\mu_1)^2}{2\sigma_q^2}}.
\]
For this Gaussian variate, the normalization is known and equal to 
\[
Z_1 = \frac{1}{\sqrt{4 \pi \sigma_q^2}}.
\]
 
\begin{figure}
\begin{center}
\includegraphics[scale=0.5]{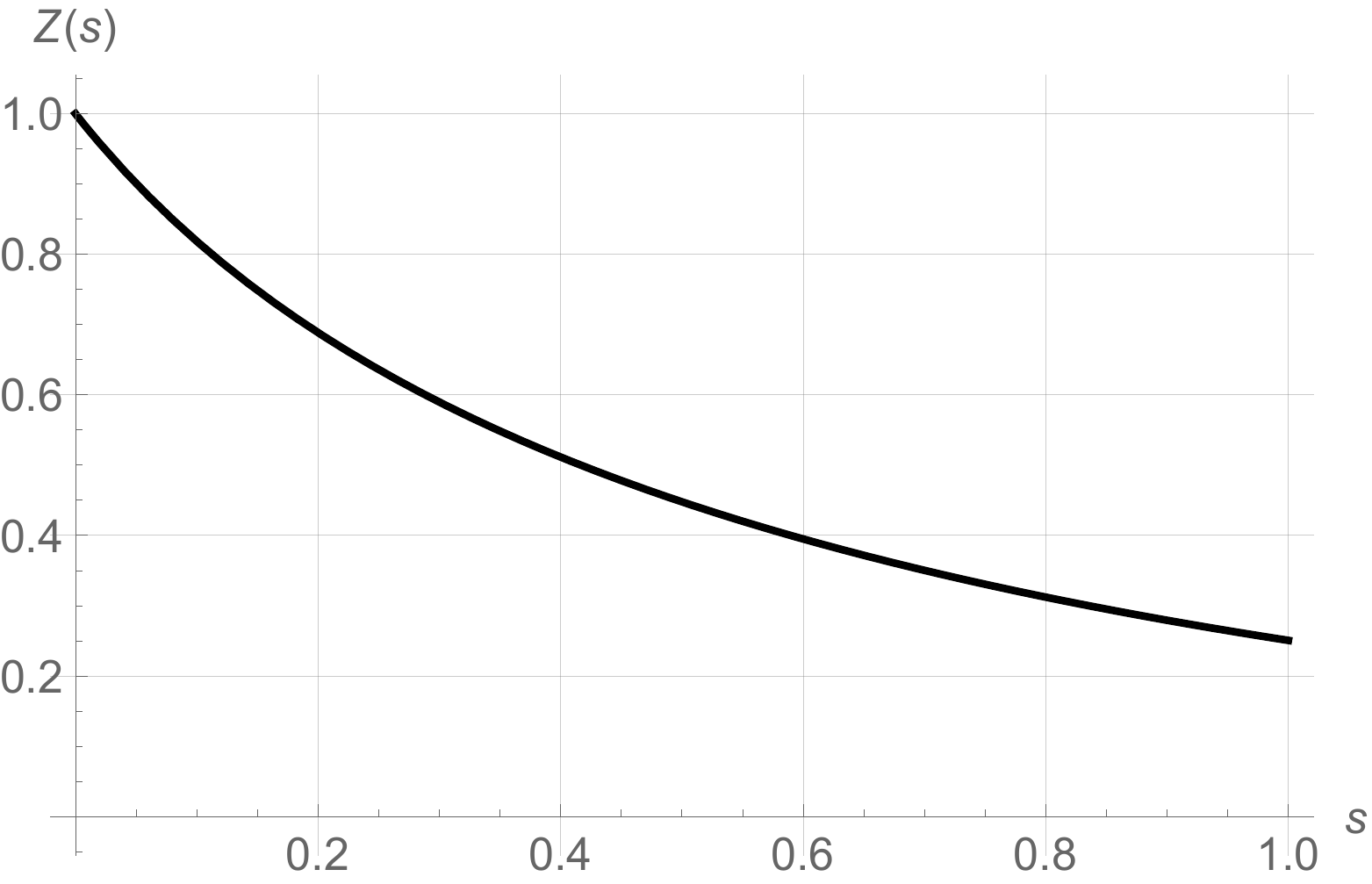}
\end{center}
\caption{Plot of $Z(s)$, with $\mu_q = \mu_p = 0$ and $\sigma_q = 0.1$ and $\sigma_p = 0.2$. See (\ref{zssimple}).}
\label{fg:zt}
\end{figure}

We will examine how the homotopy process proceeds from a starting density
\[
p=\frac{1}{\sqrt{4 \pi \sigma_p^2}} e^{-\frac{(x-\mu_p)^2}{2\sigma_p^2}},
\]
hence, $Z_0 = 1$, and thus
	\[
	\frac{Z_1}{Z_0} = \frac{1}{\sqrt{ 4 \pi \sigma_q^2}}.
	\]
In this particular example  the full path,
$Z_s$, for $0 \le s \le 1$ can be found analytically:  
	\begin{equation}
	Z_s=\int_{-\infty}^\infty q^s(x)p^{(1-s)}(x) dx =
	\frac{2^s \pi^{s/2} \sigma_p^{s-1} \sigma_q\sigma_p}{\sqrt{t\sigma_p^2+(s-1)\sigma_q^2}}
	\exp\bigg(\frac{1}{4} \frac{(\mu_q-\mu_p)^2(s-1)s}
	{\big( s \sigma_p^2+ (s-1)\sigma_q^2\big)}\bigg).
	\label{zssimple}
	\end{equation}
	 Figure \ref{fg:zt} depicts $Z(s)$ 
for the case $\mu_q = \mu_p = 0$.
For completeness we describe how the path would be approximated using sample averages via (\ref{Eq_lnZ1Z0}). In order to find  samples $X(n)_m$ we would use an efficient sample generator. However, here we can compute the cumulative density function (cdf) $F_s$ corresponding to the pdf $\theta_s/Z_s$, and further invert to find the samples at each stage $m$. Samples are obtained by computing
\[
F_s^{-1}(U) = \mu_q + (2 \sigma_p \sigma_q)^2 [\mbox{erf}^{-1}(2 U -1)]^2 + \frac{\sigma_q^2 (1-s) (\mu_p-\mu_q)}{s \sigma_p^2 +(1-s) \sigma_q^2}.
\]
where $U$ is uniformly distributed on $[0,1]$. This analytically tractable example will be used as a benchmark in Section \ref{sec.error}.

\section{Homotopy Schedule  Error}
\label{sec.error}
The error in finding $Z_1$ via the homotopy schedule is associated with approximating 
$\mu_{s}$ by the average $\bar \mu_{s}$. In what follows we derive an estimate of the error incurred in the replacement. We concentrate in the case $h(s)=s$ for simplicity, and assume $N$ is large and thus we can apply the Central Limit Theorem
to estimate the variance  associated with (\ref{Eq_scheduleZs}) and (\ref{Def_Ztilde}). Fix $\epsilon > 0$ and let $s\in [0,1]$. For large $N$ the approximate distribution of
\begin{equation}
	\bar{\mu}_s-\mu_s
	\stackrel{d}{\approx} \mathcal{N}(0, \tfrac{1}{N} \sigma^2_s), \label{Eq_error}
\end{equation}
normal with mean zero and variance $\sigma^2_s/N$ where
\begin{align}
	\sigma_s^2 &= \int \left( \frac{q}{p}\right)^{2 \epsilon} \theta_s dx - \mu_s^2 \nonumber \\
	&= \frac{Z_{s+\epsilon}}{Z_s} \mu_{s+\epsilon} - \mu_s^2
	\nonumber \\
	&= \mu_s[\mu_{s+\epsilon} - \mu_s].
\end{align}

A convenient metric to compare the true vs the estimated value of each $ Z_s $ is
\begin{equation}
	L_s := \log \left(\frac{\bar{Z}_s}{Z_s}\right)
\end{equation}
Recall $Z_{s+\epsilon} = Z_s \mu_s$ and that the corresponding recurrence holds for the estimates. Then, up to order one by a Taylor approximation,
\begin{align*}
	L_{s+\epsilon} &= 
	L_s + \log \frac{\bar{\mu}_s}{\mu_s}\\
	&\approx L_s +  \frac{\bar{\mu}_s - \mu_s}{\mu_s} + O\left( \frac{\bar{\mu}_s - \mu_s}{\mu_s}\right)^2
\end{align*}
By the calculation above, and assuming independence of $\bar{Z}_s$ and 
$\bar{\mu}_s$, we can estimate the variance as
\begin{align}
	\text{Var} \left(L_{s+\epsilon}\right) &\approx 
	\text{Var} \left(L_s \right) + \frac{\sigma_s^2}{N \mu_s^2 }   \nonumber \\
	&= \text{Var} \left(L_s\right) + \frac{1}{N} \left(
	\frac{\mu_{s+\epsilon}}{\mu_s}-1 \right). \label{Eq_estVarlog1}
\end{align}

\begin{figure}
	\begin{center}
		\includegraphics[scale=0.5]{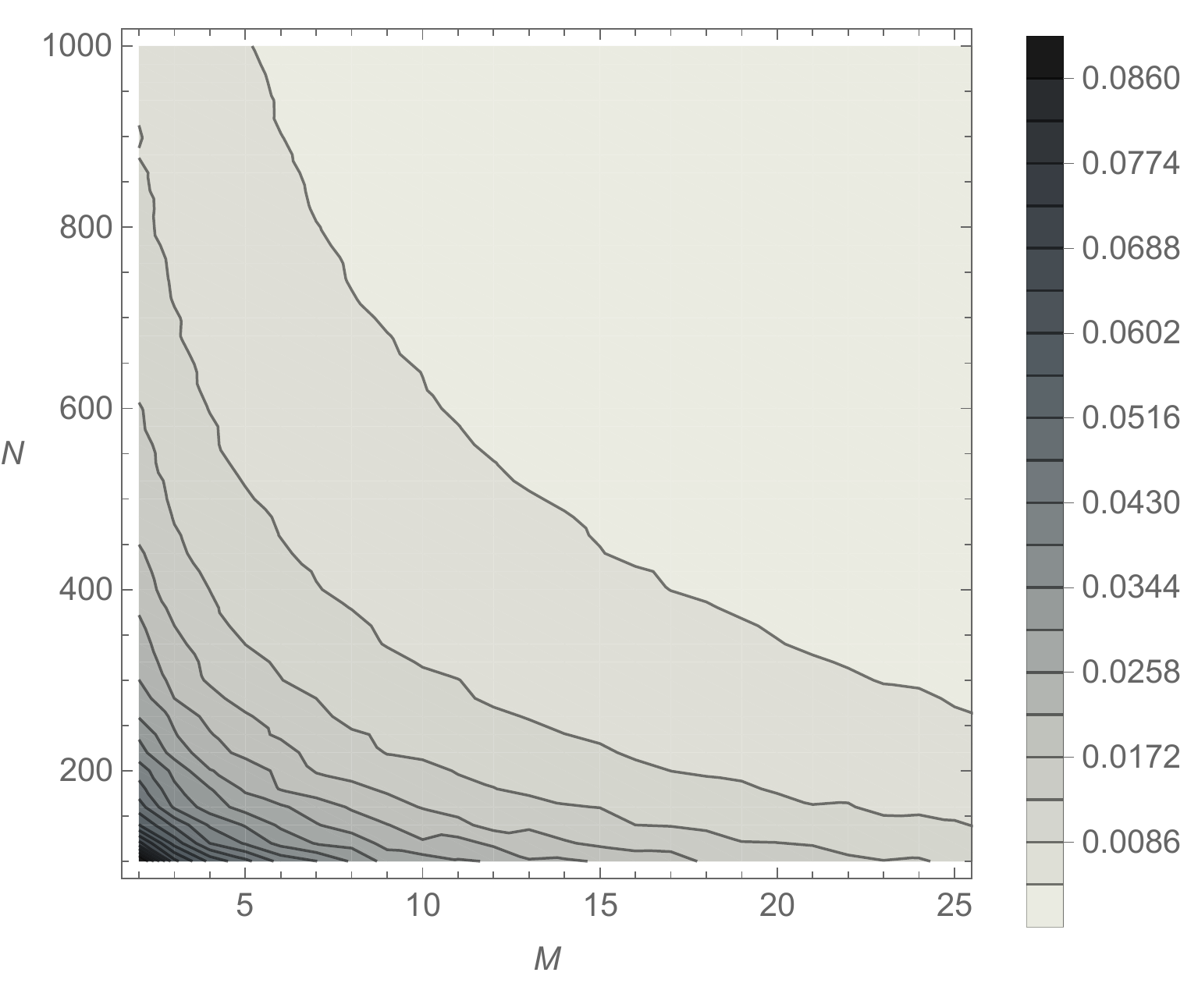}
	\end{center}
	\caption{Plot of the homotopy schedule error, as a function of $M$ and $N$, based upon the estimate in (\ref{errorhomotopy}).  Sample averages are used. The case plotted is for the estimate of $Z_1$ considered in (\ref{zssimple}), with $\mu_q = \mu_p = 0$ and $\sigma_q = 0.1$ and $\sigma_p = 0.2$. See Section \ref{example}.}
	\label{fg:error}
\end{figure}

Over the whole homotopy schedule, using $M$ equally spaced stages in $s$, and again, assuming independence, on gets the following estimate
\begin{equation}
	\text{Var}\left(\log \frac{\bar{Z_1}}{Z_1}\right) \approx \frac{1}{N} \sum_{m=2}^{M} \frac{\mu_{s_m}(\epsilon)}{\mu_{s_{m-1}}(\epsilon)} - \frac{M}{N},
	\label{errorhomotopy}
\end{equation}
where $M \epsilon = 1$. Figure \ref{fg:error} shows a plot of (\ref{errorhomotopy}) for the
calculation in Section \ref{example}, as a function of $M$ and $N$.
This figure reflects a nearly self-similar structure in the error as given by the estimate in (\ref{errorhomotopy}). We will return to this figure later on, to suggest how the inter-relation between the sample parameter $N$ and the homotopy stages $M$ can be exploited to reduce the computational complexity of the schedule in
practical applications, wherein estimates of the normalization constant or samples of the target distributions are sought.

\subsection{Generalization of the Homotopy Schedule} \label{sec.general}
A more general  homotopy formulation  that retains an inherently recursive structure 
and makes the schedule presented previously a special case is proposed next:
Let $h:[0,1]\to \mathbb{R}$ be a smooth, increasing function with $h(0)=0, h(1)=1$, and consider the following family of probability density functions
\begin{equation}
    \theta_{h(s)} = \frac{q^{h(s)}p^{1-h(s)}}{Z_s},\quad Z_{h(s)} = \int q^{h(s)}p^{1-h(s)} dx.
\end{equation}
With $\epsilon > 0$ let
\begin{equation}
    \Delta h(s) := h(s+\epsilon) -h (s),
    \end{equation} 
and define
\begin{equation}
     w_{\Delta h(s)}(x) := \left(\frac{q(x)}{p(x)}\right)^{\Delta h(s)} \label{Def_w}.
\end{equation}
Then the corresponding recursion is
\begin{equation}\label{Eq_schedule}
    \theta_{h(s+\epsilon)}(x) = \frac{Z_{h(s)}}{Z_{h(s+\epsilon)}} w_{\Delta h(s)}(x) \, \theta_{h(s)}(x).
\end{equation}

For some integrable  function $g$ 
 multiplying \eqref{Eq_schedule} by $g$ and integrating, yields
\begin{equation}\label{Eq_maing}
    \langle g \rangle_{s+\epsilon} = \frac{Z_{h(s)}}{Z_{h(s+\epsilon)}} \langle w_{\Delta h(s)} g\rangle_s.
\end{equation}
When $g \equiv 1$, we get the recursive relation associated with $\Delta h$
\begin{equation}\label{Eq_scheduleZs}
    Z_{h(s+\epsilon)} = Z_{h(s)} \, \langle w_{\Delta h(s)} \rangle_s.
\end{equation} 
When $\Delta h = \epsilon$, we recover the recurrence relationship (\ref{zzs}).
The generalized homotopy schedule counterpart  to (\ref{Eq_forwardDis}) can be written in terms of partition $0=s_0<s_1<\dots <s_M = 1$ and using sample averages as
\begin{equation}
    \bar{Z}_{h(s_{m+1})}  = \bar{Z}_{h(s_m)} \, \bar{\mu}_{s_m},
     \label{Def_Ztilde}
\end{equation}
with $ \bar{Z}_0=Z_0$ known. Here 
\[
\bar{\mu}_{s_m} = \frac{1}{N} \sum_{n=1}^{N}  w_{\Delta h(s_m)}(X(n)_{h(s_m)})
\]
and
where  $X(n)_{h(s_m)}$ is the $n^{th}$ sample from the  $(s_m)^{th}$ distribution
$\theta_{h(s_m)}$. Examples using homotopies $h$ different than the identity are presented in Section \ref{sec.sampler}.

\subsection{A Sequential Importance-Rejection Algorithm}
\label{sec.sampler}

The structure of the homotopy is used as an inspiration for a proposed sampler based upon 
the classical rejection sampling algorithm. Specifically, the proposed algorithm  follows from noting that in \eqref{Eq_schedule}, $\theta_{s+\epsilon}$  is written in a form that resembles a generalized rejection sampling method (see \cite{JSliu}). The goal of the sampling procedure is to generate samples $\mathbf{X}_{s}=\{X(i)_{s}\}_{i=1}^N$ from $\theta_{h(s)}$ to be used within the homotopy schedule.
Let 
\begin{equation*}\label{Eq_rejectionPsi}
    k = \sup_{x \in \mathbb{R}}\frac{q(x)}{p(x)}, \quad 
    \psi_{s,\epsilon}(x) =  \left(\frac{q(x)}{k p(x)}\right)^{\Delta h(s,\epsilon)}. 
\end{equation*}
If $U \sim \text{Unif}(0,1)$ and $X \sim \theta_s$, then $\theta_{s+\epsilon}$ is the conditional distribution of $X$ given $U \leq \psi_{s,\epsilon}(X)$.
Note that in the case $h(s)=s$, the rejection functions does not depend on $s$ anymore:
\begin{equation*}
     \psi_{s,\epsilon}(x) =  \left(\frac{q(x)}{k p(x)}\right)^{\epsilon}.
\end{equation*}

The importance-rejection sampler  results when it is made  sequential via the homotopy schedule. Given $N_s$ samples from $\theta_s$, generate $N_s$ realizations of $U$. Use $\psi_{s,\epsilon}$ to reject and obtain $N_{s+\epsilon} \leq N_s$ samples from $\theta_{s+\epsilon}$. Now use these samples to obtain $N_{s+2\epsilon}$ samples from $\theta_{s+2 \epsilon}$, and so on.
\medskip

\subsubsection{Numerical Example}
\begin{figure}
    \centering
    \includegraphics[scale=0.65]{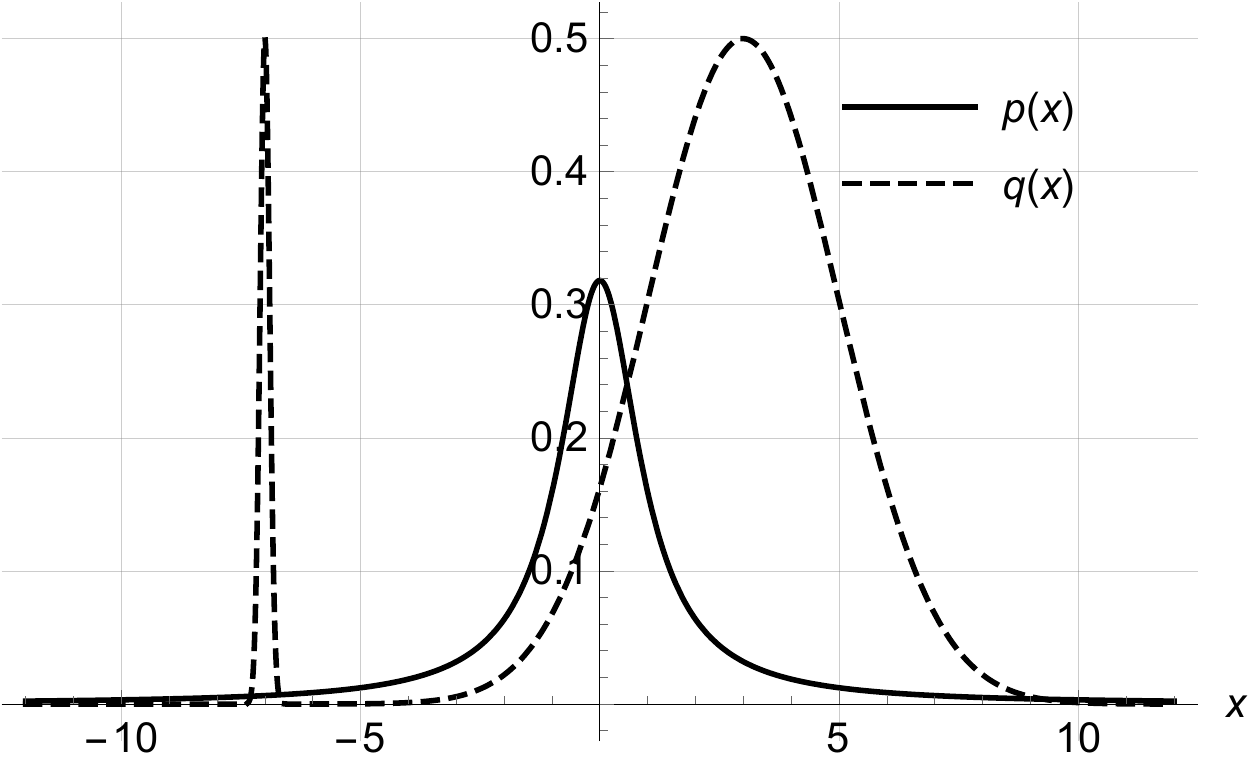}
    \caption{  Target (un-normalized) function $q$ and importance density $p$}
    \label{fig:pq}
\end{figure}

\begin{figure}
    \centering
    \includegraphics[scale=0.5]{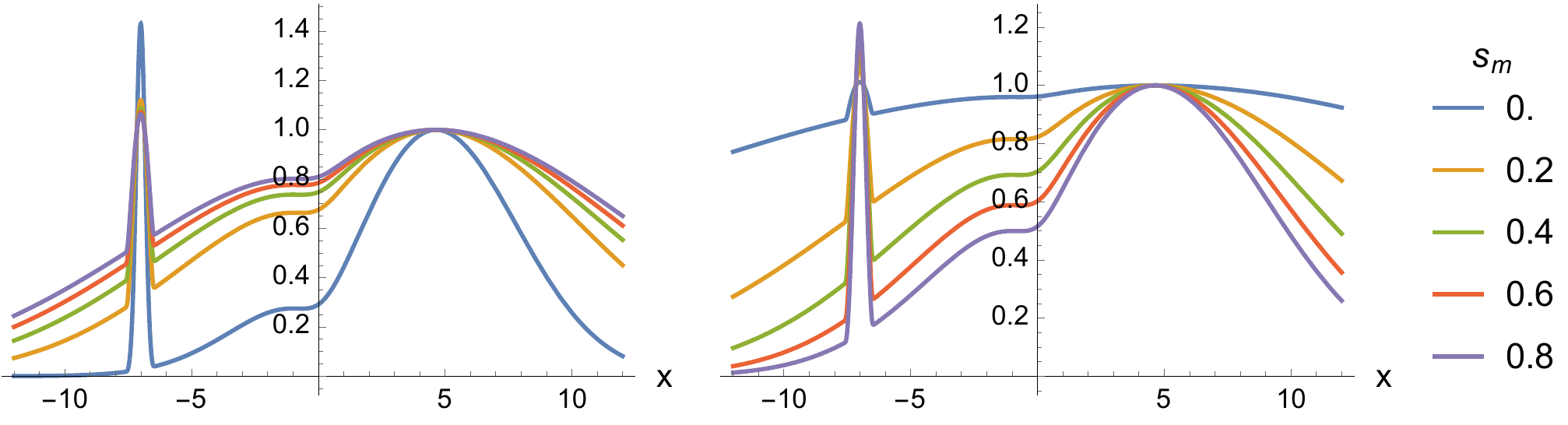}
    \caption{  Comparison of the rejection function $\psi_{s_m,\epsilon}$ of \eqref{Eq_rejectionPsi} with $\epsilon=0.2$ and different values of $m$ for $h(s) = s^{0.5}$ and $h(s) = s^{2}$.}
    \label{fig:psis}
\end{figure}

Consider a a bi-modal function $q$ composed two Gaussians, one of them with a very narrow peak, for which we want to compute the normalization constant $Z_1$. The importance probability density $p$ must be chosen such that \eqref{Hyp_pq} holds.  For a target improper distribution $q$ with exponentially vanishing support, a safe choice for $p$ is a heavy-tailed function with finite variance. Here, we take $p$ to be the density of a Student T distribution.  Figure \ref{fig:pq} shows a compares $p$ and $q$ (note that $q$ is not normalized, but $p$ is).
We present, in what follows, the results obtained for $h(s)=s^\alpha$ with $\alpha > 0$, for small and large $\alpha$. See Figure \ref{fig:psis}.
For $m=0$ the sampler generates $N_0$ uniformly distributed random variables $U_i$ in $[0,1]$, out of which it rejects all that fall above the blue curve, returning $N_1$ samples from $\theta_{s_1}$. For $m=1$, $N_1$ uniforms samples are generated and, of these, some are rejected if they fall above the yellow curve, resulting in $N_2$ samples from $\theta_{s_2}$. Continuing successively, one obtains a fraction $N_M$ of samples from $\theta_{s_M} = \theta_1$. Clearly $N_0\geq N_1 \geq \cdots N_M$. We call $N_m/N_0$ the ``rejection rates."

\begin{figure}
    \centering
    \includegraphics[scale=0.5]{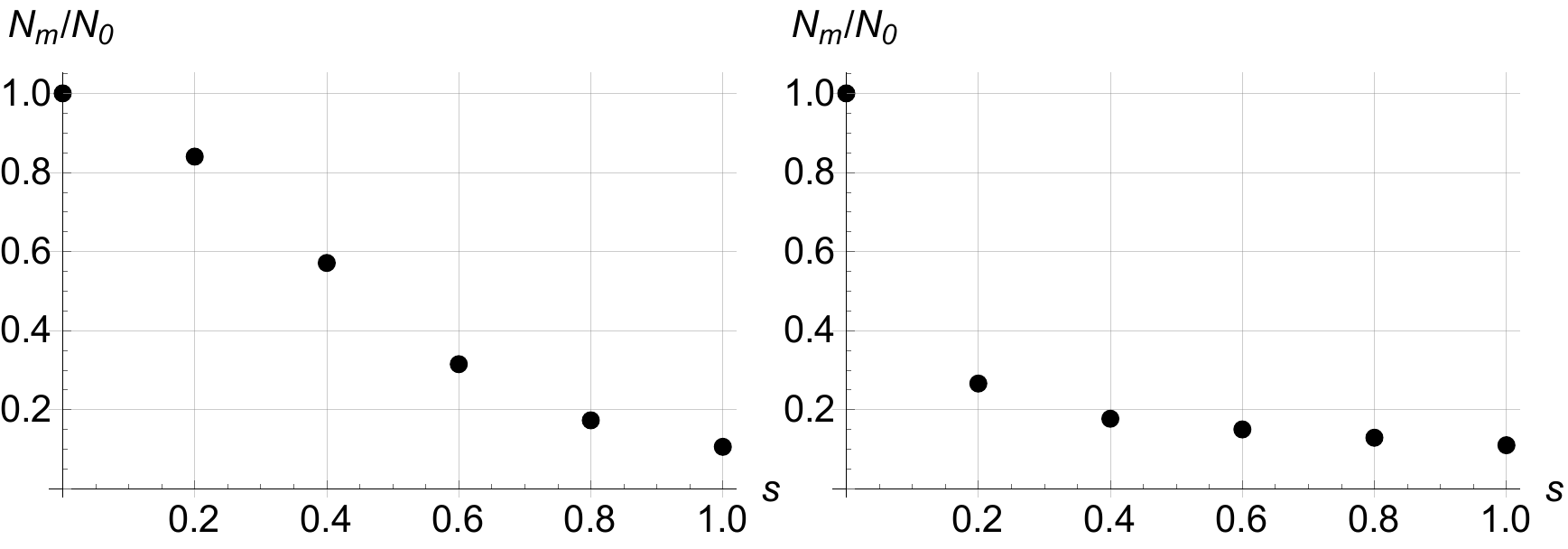}
    \caption{ Rejection rates obtained from the sequential sampler using two different homotopies. The initial sample size was $N_0=1000$. The final rejection rate is approximately $N_M/N_0 = 0.03$, and is independent of $M$ and $h$.}
    \label{fig:rejectionRates}
\end{figure}

Note that the rejection rates do not depend on $M$ but only on the choice of $h$: for $h(s)=s^{0.5}$ all rejections essentially happen for $s=0$, while in the case of $h(s)=s^2$, the rejection region becomes smaller as $m$ increases. The final rejection rate $N_M/N_0$ depends only on the choice of $p$. This is  situation is illustrated in Figure \ref{fig:rejectionRates}.

\begin{figure}
    \centering
    \includegraphics[scale=0.5]{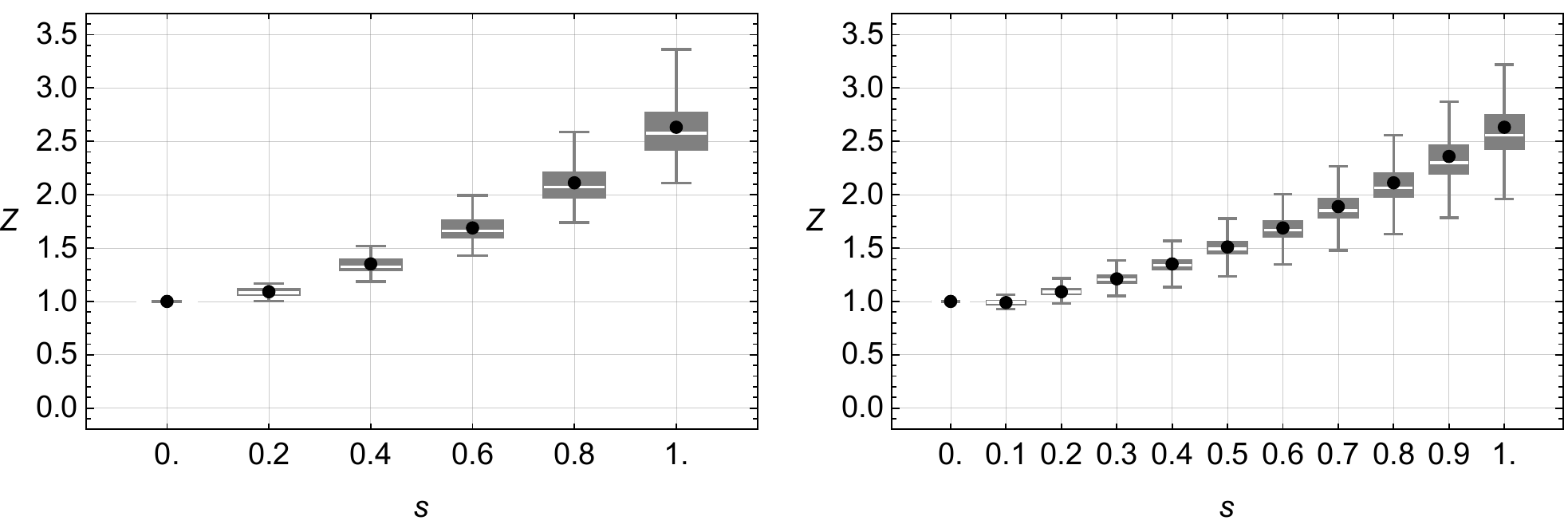}
    \caption{\small Estimates $\tilde{Z}_{s_m}$ shown as box plots from 100 realizations of formula \eqref{Def_Ztilde} using $N_0 = 1000$, $h(s)=s^2$ and different number of homotopy steps: $M=5$ on the left, and $M=10$ on the right. Red points indicate the value of $Z_{s_m}$ obtained by numerical integration of $q^{h(s)} p^{1-h(s)}$.}
    \label{fig:Zs}
\end{figure}

The normalization constant $Z_1$ can be estimated by means of the recurrence relation \eqref{Def_Ztilde}.  Figure \ref{fig:Zs} shows the boxplots for estimates of $Z_s$ in the case of $h(s)=s^2$ and two different homotopy schedules. 
It is easy to see using a simple adaptation of the results concerning rejection sampling to this
particular sampler is unbiased. Its practical value, however, is not so obvious, particularly when computational efficiency is important. Other sampling algorithms, such as the systematic resampling algorithm  \cite{wolresampling}, could be used in the homotopy schedule when 
efficiency is a primary concern.

%
%
%

\section{Computational Complexity}
\label{sec.complexity}

\begin{figure}
\begin{center}
\includegraphics[scale=0.5]{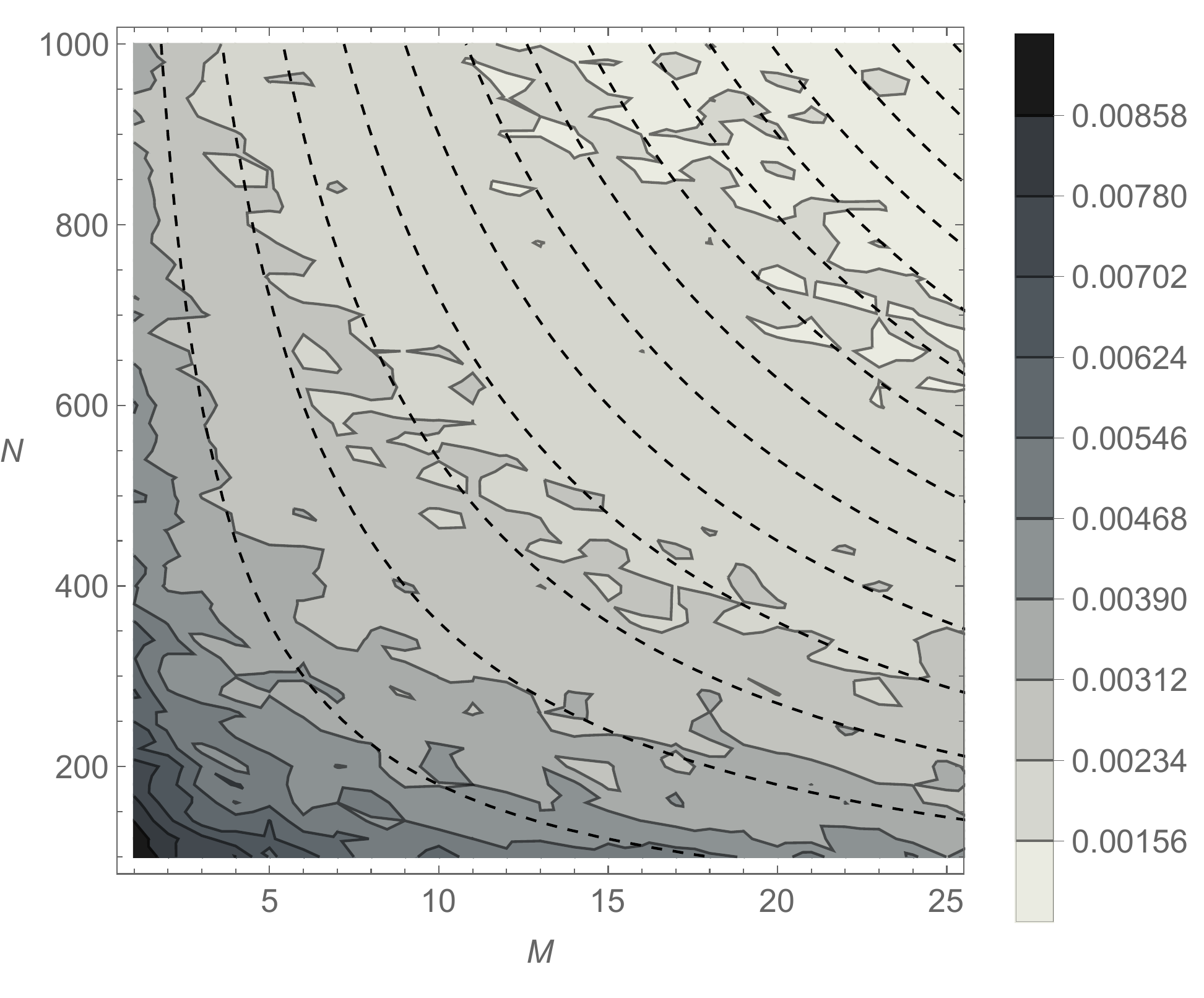}
\end{center}
\caption{For the Gaussian example in Section \ref{example}: dependence of the error $ |\tilde Z(M,N)-Z_1|$ on homotopy steps $M$ and number of samples $N$ per step. Here  $\bar Z_1(M,N)$ is the computed estimate of $Z_1$. For each combination of $(M,N)$, the error is computed as the average over 100 runs. Constant $M \times N$  contours are superimposed.}
\label{multig}
\end{figure}

The computational cost  of the homotopy schedule is proportional to $M \times N$. 
Figure \ref{multig} is a contour plot of the error norm $ |\bar Z_1(M,N)-Z_1|$, where $\bar Z_1(M,N)$ is the computed estimate of $Z_1$ for the Gaussian case considered in Section \ref{example}. The figure shows how the error
depends on the number of homotopy steps $M$ and the number of sample averages $N$. 
The plot is the result of the average of 100 experiments for each pair $(M,N)$. 
Superimposed on the experimental outcomes are the contours of equal $M \times N$, a family of hyperbolic curves.  These numerical outcomes agree qualitatively with the estimate in (\ref{errorhomotopy})  (see Figure \ref{fg:error}). 

The homotopy accuracy has an inherently hyperbolic curve dependence on $M$ and $N$. The two figures thus suggest that, in a computation, it is
possible to exchange sample steps and the number homotopy stages. This characteristic can be exploited in the computational setting on hybrid  CPU/GPU computing architectures.

\section{Homotopy Schedules on Bayesian  Stationary Processes}
\label{sec.bayes}

We propose to use the homotopy schedule to 
 compute the {\it evidence}
\[
Z_1 = \int \pi(d|x) \pi (x) dx :=\int q(x)  \, dx,
\]
where  $\pi(d|x)$ and $\pi(x)$ are the likelihood and the prior, respectively.

Several alternative homotopy schedules arise. The first two require that the prior or the likelihood be proper distributions. In the third alternative, it is assumed that the distributions are improper and thus follows in a straightforward way from the prior sections. In the first two
we will assume that at least  the prior distribution is proper, but the modification to these two alternatives when the likelihood is proper is obvious. A useful characteristic of the homotopy schedule when either the prior or the likelihood is used for $p(x)$ is that the support of $p(x)$, {\it vis-a-vis},  the support of the posterior,  is not a concern.


For a first alternative, we want to perform a homotopy from the prior $p=\pi(x)$, at $s=0$, to $q= \pi(y|x) \pi(x)$, at $s=1$. 
We chose $p$ to be the prior assuming that this is a proper distribution, {\it i.e.}, that $Z_0$ is known for this case:
$Z_0 = \int \pi(x) dx$. Then
\begin{equation}
Z_s= \int [ \pi(y|x)]^s \pi (x) dx.
\label{zbayes1}
\end{equation}

Using the same reasoning as before, we can estimate, approximately,
\begin{equation}
\log Z_1  \approx \sum_{m=1}^M \log \frac{1}{N} \sum_{n=1}^N \pi(y \, |X(n)_{m})^\epsilon + \log Z_0,
 \label{z1}
 \end{equation}
 where  $X(n)_m$ is the $n^{th}$ sample from the $m$-th distribution, {\it i.e.}, 
 \[
 \theta_{m \epsilon}(x) = \frac{1}{Z_{m \epsilon}} \pi(x) \pi(y|x)^{m \epsilon}.
 \]
 
A second alternative: supposing we know the $Z_0$ associated with the likelihood. It is obvious then  that one can reformulate (\ref{zbayes1}) to read instead
 \begin{equation}
Z_s = \int [ \pi(x)]^s \pi (y|x) dx.
\label{zbayes2}
\end{equation}
in which case an approximation is
\begin{equation}
\log Z_1  \approx \sum_{m=1}^M \log \frac{1}{N} \sum_{n=1}^N \pi(X(n)_{m})^\epsilon + \log Z_0,
 \label{z12}
 \end{equation}
 where  $X(n)_m$ is the $n^{th}$ sample from 
\[
\theta_{m \epsilon}(x) = \frac{1}{Z_{m \epsilon}} \pi(x)^{m \epsilon} \pi(y|x).
\]
We remark here that $Z_s$ will have a different starting value, depending on whether (\ref{z1}) or (\ref{z12}) 
is used, further, it generates two different orbits $Z_s$.
 Having the flexibility of being able to use (\ref{z1}) or (\ref{z12}) can be useful when generating samples of 
 the prior or the likelihood is easier than the other way around. It could also be dictated by whether or not a starting normalization is known.


An alternative formulation arises if we  appeal to an auxiliary distribution $p=I(x)$, for which 
$Z_0 = \int I \, dx$ is known.  This is the strategy to use when the prior and the likelihood are both improper distributions. Provided the support of $I(x)$ is larger than the prior times the likelihood, 
 \begin{equation}
Z_s = \int \left[ \frac{\pi(x) \pi (y|x)}{I(x)} \right]^s  I(x) dx.
\label{zbayes6}
\end{equation}
we leave the consequent details omitted for this case as it follows in a straightforward way.

\subsection{Example Calculation}
The following example compares the homotopy schedule with the analytical 
estimation in a Bayesian application. In these $y=0.65$, $R=0.25$, and $Q=0.2$.
 
 We take
\[
\pi(x) = \frac{x}{R^2} \exp [ -x^2/2 R^2],
\]
a Rayleigh distribution. Samples of this distribution are    $X = \sqrt{-2 \log U}$, where $U$ are uniformly distributed on $[0,1]$. Suppose the posterior has the form
\[
\pi(x) \pi(y|x) = \pi(x) \exp [(x-y)^2/2 Q^2].
\]

For this case, using (\ref{z1}) 
\begin{equation}
Z_s = \frac{\pi}{2} \frac{Q R y s}{(Q^2 + R^2 s)^{3/2}} \exp 
\left[ \frac{s y^2 }{2(Q^2 + R^2 s)}
\right]
  \left(2 -    \mathcal{P} \left[-\frac12, \frac{(R^2 s^2 Y^2)}{2 (Q^4 + Q^2 R^2 s) } \right]\right),
\label{z1rayleigh}
\end{equation}
where $\mathcal{P}$ is the regularized Gamma function. When the homotopy schedule is used, with $M=10$, and $N=50$, the maximum error between the analytical and the sample average outcomes was less than $10^{-8}$.

%
%

\section{The Time Dependent Bayesian Case: Particle Filters and  Data Assimilation}
\label{sec.assimilation}

%
%
Data assimilation \cite{Ev96, sarkkabook} is the name given to a variety of different Bayesian estimation methods that 
combine noisy model outcomes and observations to produce estimates of state variables.
In the context of state estimation in time dependent problems it is analogous to filtering.
In its simplest guise, moments of the posterior distribution of the state variable $x_t$, conditioned on observations $y_t$
for an evolutionary problem are to be estimated. In the Bayesian framework the model is used to inform the prior and the observations to inform the likelihood.

Consider the time-discrete stochastic {\it model}
\begin{equation}
x_{t+h} = Q(x_t,t) + \sigma_x \Delta w_t, \quad t=0,h, 2h, ...
\label{modeleq}
\end{equation}
over the time span
$0 \le t \le T$. The initial condition $x_0$ is drawn from a known initial distribution. The $\Delta w_t$ is a Wiener process with known variance $\sigma_x$.
The
{\it observations} are given by 
\begin{equation}
y_t = H(x_t) + \sigma_y \varepsilon_t, \quad t=0,h, 2h, ...
\label{obseq}
\end{equation}
for $0 \le t \le T$. (We are assuming that the observations and the model outcomes are available at all $t$ steps, for simplicity).  $H$ is the observation function, $\sigma_y$ is the variance and $\varepsilon_t$ 
is a standard normal. The noise processes are assumed here to be uncorrelated in time and uncorrelated with each other, for simplicity.

The general goal is to find estimates of moments of $x_t$ given observations $y_t$, for 
$t\in[0,T]$, from the posterior distribution
\[
\pi(x_{t=0:T}| y_{t=0:T}) = \prod_{t=0}^T \frac{1}{Z^{(t)}}  \pi(y_{t=0:T}|x_{t=0:T})  \pi (x_{t=0:T}).
\]
(A common estimator is the minimizer of the posterior covariance). The quantity $Z^{(t)}$ is generally not known. Particle filters (see \cite{sarkkabook}) are often used when the dynamics are nonlinear, and in general do not require knowledge of the normalization factor $Z^{(t)}$. However, it is well-known that the filter will `collapse' if 
the number of particles used is small. Resampling is commonly employed to counter this problem. However, resampling is practically unavoidable in finite-precision, finite-resource computing, when particle filters are applied in practice to even moderate dimensional problems and/or when nonlinear/non-Gaussian processes are involved. A number of 
different resampling strategies have been proposed  (see \cite{wolresampling}).

We will make use of the homotopy schedule to find the normalization $Z^{(t)}$, at each $t$,  as a means to address particle filter collapse. This is especially useful when  $N$ is by necessity, small.

\subsection{Data Assimilation via Homotopy Particle Filters}
\label{dapf}

The algorithm for the standard particle filter with resampling is described in \cite{sarkkabook}. The particle filter algorithm samples the prior, evolves the particles through the model, and computes weights according to the likelihood given the data. In essence, both the prior and the posterior are approximated by a discrete distribution, the latter is normalized at each step by the sum of the weights. Here we restate the basic estimation problem, and propose a method that combines discrete and continuous representations of the posterior where the explicit normalization of a pdf is needed. For that, we use homotopy.

We wish to find time dependent samples and moments of the posterior distribution 
$\pi(x_{t+h}|y_{t+h})$. Using Bayes,
\begin{align}
\pi(x_{t+h}|y_{t+h}) &= \int \pi(x_{t+h},x_t|y_{t+h}) dx_t \nonumber \\
&=\frac{1}{Z^{(t+h)}} \int \pi(y_{t+h}|x_{t+h},x_t)  \pi(x_{t+h},x_t|x_t) dx_t\nonumber \\
&=\frac{1}{Z^{(t+h)}} \int \pi(y_{t+h}|x_{t+h})  \pi(x_{t+h}|x_t) \pi(x_t) dx_t \nonumber \\
&=\frac{\pi(y_{t+h}|x_{t+h})}{Z^{(t+h)}} \int  \pi(x_{t+h}|x_t) \pi(x_t) dx_t
\label{pieq}
\end{align}
where is the $Z^{(t+h)}$ is the normalization constant. It is found via
\begin{equation}
Z^{(t+h)}= \int \pi(y_{t+h}|x_{t+h}) \left[  \int   \pi(x_{t+h}|x_t) \pi(x_t) dx_{t} \right]  dx_{t+h}.
\label{zth}
\end{equation}


Suppose one has assimilated data up to time $t$ and knows the probability density function of the prior $\pi(x_t)$. Additionally, suppose the stochastic process is such that the transition density $ \pi(x_{t+h}|x_t)$ is know (e.g. a $x$ is the solution to a stochastic differential equation with additive Gaussian noise).  Via a homotopy calculation we can obtain an approximated density for the posterior $\pi(x_{t+h}|y_{t+h})$ as follows. Let $x_t^1,...,x_t^n$ be (un-weighted) samples from $\pi(x_t)$ and approximate \eqref{zth} by
\begin{equation}
	Z^{(t+h)}\approx \int \pi(y_{t+h}|x_{t+h}) \left[ \frac{1}{n}\sum_{i=1}^{n}  \pi(x_{t+h}|x_t^i) \right]  dx_{t+h}.
	\label{zthApp}
\end{equation}

The term in brackets is obtained by advancing the model from the sampled ``particles" $x_t^i$ as in the standard particle filter, but here we keep track of the complete transition density function. Namely, as a function of $x_{t+h}$, the term in brackets is a known density (typically, an averaged sum of Gaussians) and hence can be use as a starting function $p$ for the homotopy schedule. Once an estimate $\tilde{Z}^{(t+h)}$ of the normalization constant is obtained, one gets the following approximate density function for the posterior

\begin{equation}
	\pi(x_{t+h}|y_{t+h}) \approx \frac{\pi(y_{t+h}|x_{t+h})}{\tilde{Z}^{(t+h)}} \frac{1}{n}\sum_{i=1}^{n}  \pi(x_{t+h}|x_t^i),
	\label{pieqApp}
\end{equation}
which, in turn, can be used as the prior at time $t+h$.

 From the conditional marginals $\pi(x_t,|y_t)$, it is possible to fully determine an approximation for the distribution
 $\pi(x_{t=0:T}|y_{t=0:T})$, where $T$ is the final time. Samples can then be drawn from
 $\pi(x_{t=0:T}|y_{t=0:T})$ with which to estimate the requisite moments of $x_t$ given observations $y_t$. If the dimension of $x$ and $y$ are small, this program is not unreasonable.  The nature of the homotopy schedule, however, requires that the support of the 
 starting posterior, at time step $t$ be larger than at $t+h$. This is an unreasonable constraint in the general case, unless the variance of the distribution at $t$ is artificially inflated when required.

\subsubsection{Example Calculation}

We will describe the homotopy filter procedure by applying it to a well-known 
noisy dynamics problem. We also 
compare the accuracy and computational cost of the homotopy particle filter, with the standard particle filter \cite{godsill01monte}. 

The example problem is taken from \cite{kittagawa,gordonsalmondsmith}. 
In what follows we will use the term `position' to denote the random variable $x_t$. The {\it model} outcomes are given by
\begin{equation}
x_t = 0.5 x_{t-1} + 25 \frac{x_{t-1}}{(1 + x^2_{t-1})}+ 8 \cos[1.2(t-1)] + w_t, \quad t=1,2,..,T.
\label{model}
\end{equation}
The initial state is drawn from $x_0$ which is assumed to be Gaussian with zero mean and unit variance. 
The {\it observations} are
\begin{equation}
y_t = x_t^2/20 + v_t, \quad t=1,2,...,T-1,
\label{obs}
\end{equation}
where $w_t$ and $v_t$ are zero mean Gaussian processes of unit variance.  The time step is $\Delta t = 0.5$, which we call the filtering time steps. In the computations that follow, we will be reading  observations at every time step, hence the filtering time step coincides with the model time step. 

We will highlight an example calculation that demonstrates numerically that  the homotopy particle filter can produce results that are superior to those obtained with  a standard particle filter. We focus on a case when very few particles are used in the estimate.

Prior to examining how the homotopy particle filter performs as an estimator in the dynamics of the problem, we can focus on how well it was able to estimate the normalization constant of the posterior distribution.
We tracked the error in computing the estimate of the normalization constant \eqref{zth} of the posterior via the homotopy particle filter and a quadrature-based `truth' estimate. It was found that the ratio of the homotopy-derived constant to the quadrature-derived constant, over the whole time span, had mean value of 1.04 and a standard deviation of 0.13. The error in this ratio was more pronounced at times when the particle distribution was fairly concentrated, as compared with the support of the pdf distribution. However, a striking feature of the homotopy particle filter estimate was that the discrepancy did not lead to catastrophic collapse of the homotopy particle filter. This is notable since the standard particle filter, particularly in the small particle cases, can develop poor particle distributions which even with resampling, cannot avoid collapse.

\begin{figure}
\begin{center}
\includegraphics[scale=0.4]{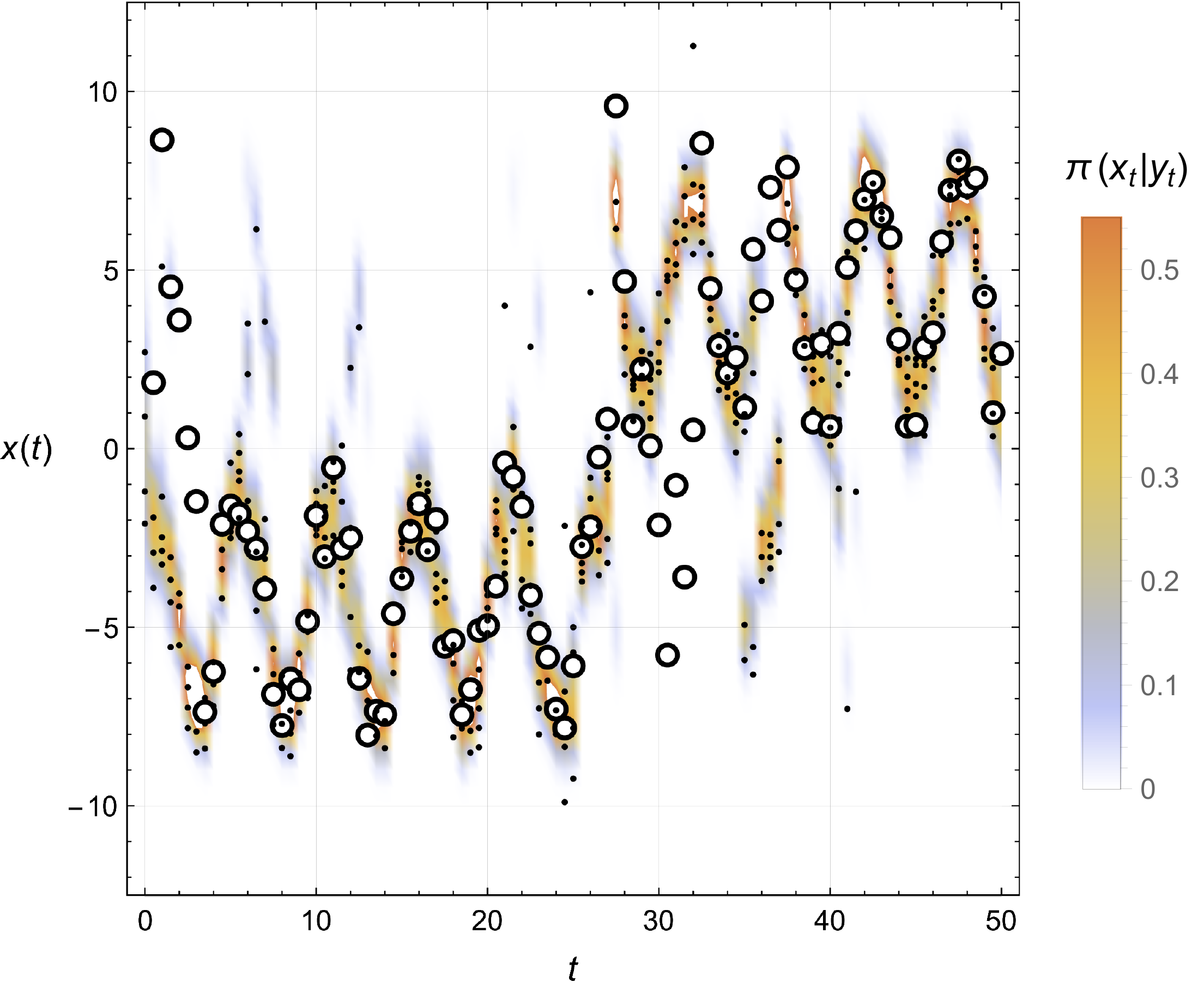}
\end{center}
\caption{Plot of the pdf of the posterior distribution of the position, conditioned on observations associated with (\ref{model})-(\ref{obs})  problem,  and approximated via the homotopy filter. The  circles represent the estimated  mean $x_t$ from the homotopy particle filter. Black dots indicate the sampled values of $\pi(x_t)$ ($N=10$ particles were used at every step). The variance on the model and the observations was 1. The filtering and model time steps are both $\Delta t = 0.5$.}
\label{fg:particlea}
\end{figure}

The homotopy particle filter produces the marginal densities $\pi(x_t|y_t)$ conditional on the data up to time $t$,  which then can be combined to compute samples and moments from the path. We will see  in the numerical examples that follow , that the homotopy particle filter produces marginals that concentrate mass around the actual values of $x_t$.
Figure \ref{fg:particlea} shows the evolution of the conditional marginals $\pi(x_t|y_t)$ in color, compared to the model outcomes $x_t$ in circles. The black dots indicate the samples $x_t^i$ taken from the prior $\pi(x_t)$ at each step.  The number of samples used was $N=10$.
The plot suggests that the 
homotopy particle filter is a robust estimator in this problem, even when the number of particles used is small. The results shown are consistent with simulations wherein a higher number of particles were used.

\begin{figure}
\begin{center}
(a)\includegraphics[scale=0.4]{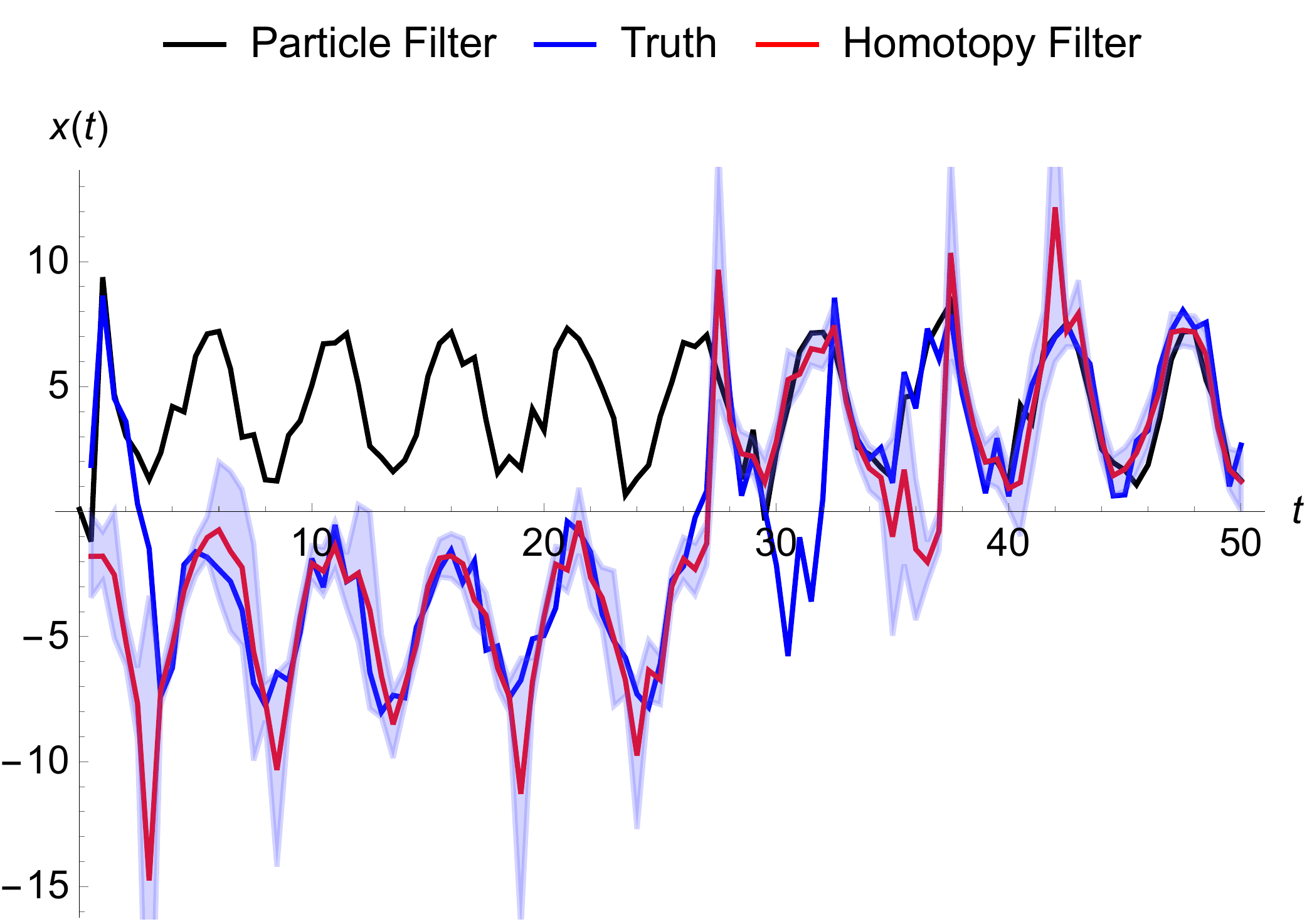}\\
(b)\includegraphics[scale=0.4]{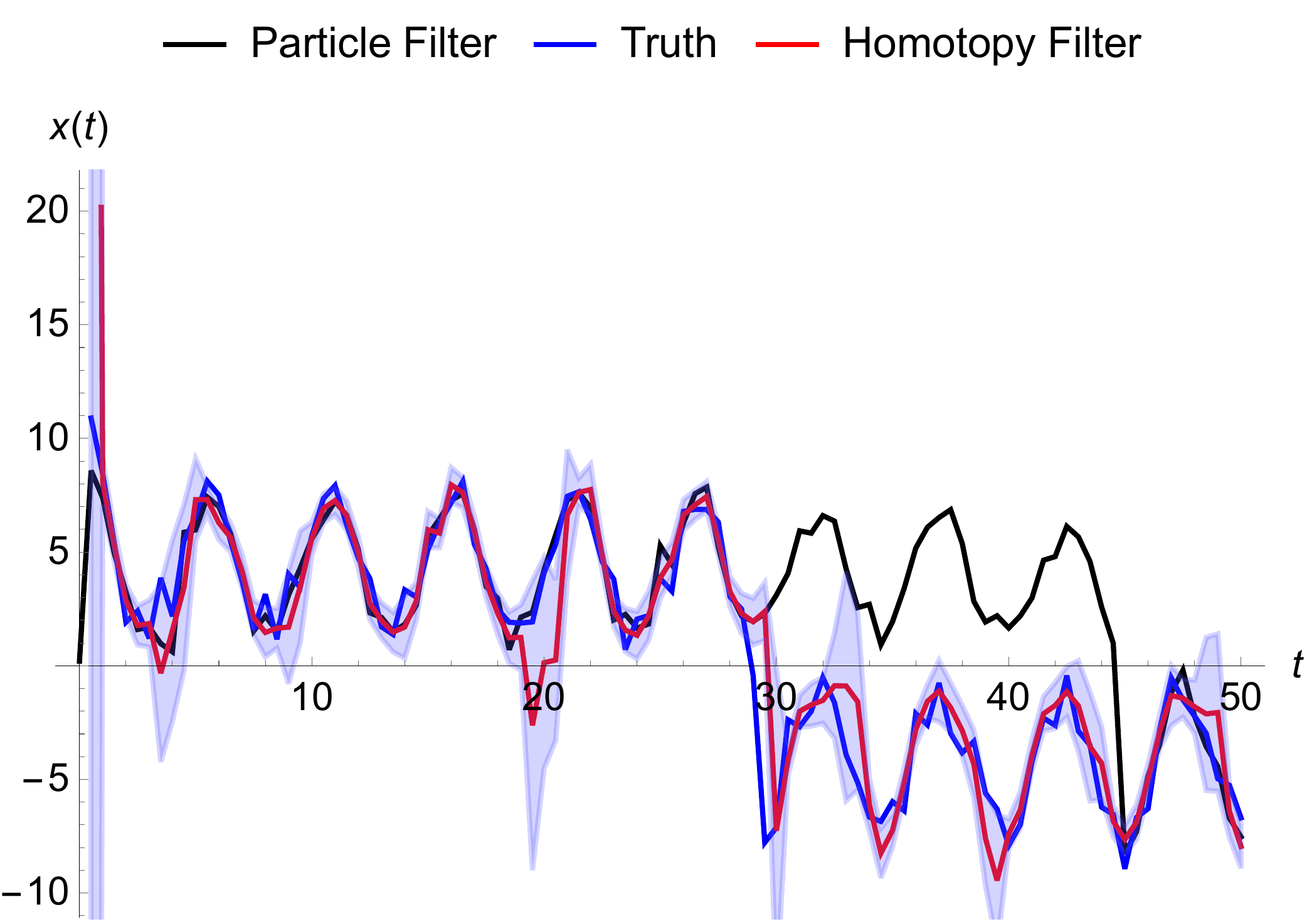}
\end{center}
\caption{ Filter estimates of the mean for the posterior position given observations associated with  problem (\ref{model})-(\ref{obs}). The filter time steps and time steps are $0.5$, the variance on the model and the observations are set to unity.
Shown is a comparison of the true path $x_t$ (blue), the mean from the standard particle filter (black) and the mean from the homotopy particle filter (red). 
The uncertainty around the homotopy particle filter mean is shaded in purple.
(a) $N=5$, (b) $N=10$.}
\label{fg:particle1}
\end{figure}

In Figure  \ref{fg:particle1}a we show comparisons of the true path, the standard particle filter posterior mean, and the homotopy particle filter posterior mean for the same setup and realization as in Figure \ref{fg:particlea}. The homotopy particle filter and the standard particle filter are executed using the same number of particles, $N=10$.
 The true path is generated by running the model (\ref{model}), starting from a sample from the  position initial distribution. The true path appears in (dark) blue in the figures. The standard particle filter mean position is shown in black. The red curve corresponds to the mean position computed at each time from the marginal $\pi(x_t|y_t)$ obtained by the homotopy particle filter.  In light blue we show the uncertainty region, defined as the region bounded by 1 standard deviation away from the estimated mean.  
  
 In Figure \ref{fg:particle1}b the experiment is repeated with $N=5$ particles.  The homotopy particle filter is able to better track the truth better than the standard filter, even when the particle density is  small. The results shown in these figures are typical of a large number of runs, for this particular problem. Though not shown, there were numerical runs that resulted in  a successful estimation outcome from the homotopy particle filter and a failed standard particle outcome. 

\section{Conclusions}
\label{sec.conclusions}

A sampling procedure is proposed, aimed at estimating the normalization constant 
$Z_1 = \int q(x) dx$ of a positive function $q$ in order to arrive at a probability density function $q/Z_1$. The homotopy is performed on the function $Z(s)$, for $s\in [0,1]$, which takes on a known value $Z_0$, associated with the distribution $p(x)/Z_0$.  In addition to the requirement that $p(x)/Z_0$ be known, it is also necessary that the support of $p(x)$ be larger than the target  $q(x)$. As $s$ is varied from 0 to 1, samples of the homotopy distribution $\theta_s(x)=  \frac{q^s(x)p^{(1-s)}(x)}{Z_s}$ are generated. These are used within an iterative scheme to generate the desired estimate, $Z_1$. A discretization in $s$ of the iteration process generates a numerical algorithm
with which to carry out the calculation numerically. We call this algorithm the homotopy schedule.

A potential application of the homotopy schedule  is in the estimation of the microcanonical ensemble in statistical physics applications \cite{vankampenbook}.

A feature of the schedule is that one can exchange the computational expense of sampling 
with the number of stages in the schedule, while keeping the error of the outcomes approximately equal. This exchange can be exploited to improve the efficiency of the calculation on a given particular computer architecture.

The homotopy schedule was  extended to Bayesian target distributions. In the Bayesian setting the requirement  on the extent of the support of $p(x)$ disappears if either the prior or the likelihood are proper distributions and used as the starting distribution for the schedule.

\par
A potential application of the homotopy schedule is in the calculation of the  normalization constant associated with filtering procedures in Bayesian estimation  for time dependent problems. These estimation problems are also known as data assimilation problems. We focus on particle filter data assimilation. In high dimensional problems it is often the case, in applying particle filters,  that the number of particles required in the estimation procedure is large in comparison to the dimensions of the state variable, in order to avoid filter collapse. The requirement of a large number of particles can prove to be challenging to achieve in practice.  We showed that the homotopy schedule could prove useful  in taming filter collapse in Bayesian estimation that make use of particle filters when the particle number of small.  

As discussed in connection with applying the homotopy schedule on stationary Bayesian estimation, in time dependent Bayesian estimation problems, there are a number of different ways to pose the homotopy if either the likelihood and/or the prior proper distribution is known. When either of these distributions is used as a starting distribution (presuming the one chosen is a proper distribution), the schedule's requirement on the  support of the target posterior and the starting distribution becomes irrelevant. This aspect of the homotopy schedule, as applied to the particle filters, was not fully explored here but could be exploited in the practical setting.

Homotopy has been applied in data assimilation previously, however, it was proposed as an alternative method for variance inflation (see \cite{even97}): 
In  \cite{maroulas} the authors apply homotopy to the  data assimilation  process as a way to handle nonlinearity and its non-Gaussian consequences on the  target posterior distribution
 of the state variables, conditioned on observations.They propose to modify the particle filter by replacing (\ref{modeleq}) by
  \begin{equation}
x_{t+h} = s Q(x_t,t)  (s-1) P(x_t,t) + \sigma_x \Delta w_t, \quad t=0,h, 2h, ...
\label{modelmaroulas}
\end{equation}
where $s\in[0,1]$. The drift term $P(x_t,t)$ is thus used as a 'steering' term, the target model
has no $P$ and a drift $Q$. The degree to which steering takes place is then controlled by $s$.

The homotopy schedule may prove useful in other data assimilation strategies.
The ensemble Kalman filter (enKF) is a popular data assimilation technique \cite{Ev96}. It is 
based on the Kalman Filter and was largely motivated by finding an alternative to 
the extended Kalman Filter, to do Bayesian estimation on problems, such as  (\ref{modeleq})-(\ref{obseq}).  In the enKF the `prediction' step consists of drawing samples from the 
distribution $\pi(x_{t}|y_t)$, and advancing these to time $t+h$. Using a Gaussian assumption
the `analysis' step consists of updating the ensemble to assimilate observations at $t+h$.
In \cite{youssefmarzouk} the authors propose using the Rosenblatt rearrangement \cite{rosenblatt} to forgo the Gaussian assumption implicit in the analysis step in the enKF.
It could  be argued that knowing the normalization constant of the marginals could lead to more informed Rosenblatt rearrangements, hence, a combination of both techniques may prove useful, computationally. 

Another data assimilation methodology is the path integral method \cite{path, drifter}.
The path integral formulation is capable of 
handling very general nonlinear dynamics and non-Gaussian statistics, but it is severely challenges with regard to the dimension of the state space. The homotopy procedure can also prove useful in estimating 
the normalization constant of the posterior.   If the state estimation problem is low dimensional, in terms of the dimension of the state variable, the homotopy estimation of the  normalization constant for the posterior over the whole time trajectory will produce an approximation to the 
full posterior distribution. Such a detailed probabilistic description of the dynamics can then be sampled directly to obtain derived estimates of the time dependent moments of the state, conditioned on observations.


\section*{Aknowledgments}
The submitted manuscript has been authored by a contractor of the U.S. Government under Contract No. DE-AC05-00OR22725. Accordingly, the U.S. Government retains a non-exclusive, royalty-free license to publish or reproduce the published form of this contribution, or allow others to do so, for U.S. Government purposes.
This work was also  supported by  and by NSF DMS  grant 0304890 and NSF OCE grant 1434198. Part of this work was carried out at NERSC, in Bergen Norway, and at
 Stockholm University through  its Rossby Fellowship Program.











\bibliographystyle{plainnat}

\bibliography{doubwell}
\end{document}